# A feasible roadmap for developing volumetric probability atlas of localized prostate cancer

Liang Zhao[1,2], Jianhua Xuan[1], and Yue Wang[1,*]

[1]*Department of Electrical and Computer Engineering, Virginia Polytechnic Institute and State University, 900 N. Glebe Road, Arlington, VA 22203, USA*
[2]*School of Information and Control Engineering, Xi'an University of Architecture and Technology, 13 Yanta Road, Xi'an, Shaanxi 710055, China*
[*]*yuewang@vt.edu*

**Abstract:** A statistical volumetric model, showing the probability map of localized prostate cancer within the host anatomical structure, has been developed from 90 optically-imaged surgical specimens. This master model permits an accurate characterization of prostate cancer distribution patterns and an atlas-informed biopsy sampling strategy. The model is constructed by mapping individual prostate models onto a site model, together with localized tumors. An accurate multi-object non-rigid warping scheme is developed based on a mixture of principal-axis registrations. We report our evaluation and pilot studies on the effectiveness of the method and its application to optimizing needle biopsy strategies.

## 1. Introduction

Prostate cancer is the most prevalent male malignancy and second leading cause of cancer death in men. Though most prostate cancers are slow growing; there are cases of aggressive prostate cancers [1]. The important tools for better outcome of clinical treatment to rescue patients with prostate cancer include early detection and personalized diagnosis [2]. At present, diagnosis of prostate cancer heavily relies on the pathological examination of stained tissue samples acquired by multiple while almost random biopsies. Experiment results show that one out of five cancers will be missed by the existing needle biopsy protocols. The unsatisfactory clinical diagnosis leads to the fact that the best clinical treatment time window may be missed because of undetected cancer while many patients with dormant tumors are over-treated.

Here we report a feasible roadmap for developing a volumetric probability atlas of localized prostate cancer using optically-imaged surgical specimens [3] and image/graphics processing methods [4]. This master model contains a precise probabilistic map of localized prostate tumor distribution and the corresponding anatomic structure of a prostate site model. Base on the developed statistical atlas and visualization technique, we can better understand the spatial distribution of prostate cancer with various grade, uncover the mechanism responsible for tumor behavior; and propose an atlas-informed biopsy sampling strategy.

The construction of a volumetric probability atlas of localized prostate cancer includes the following major steps:
(1) Raw data collection and pre-processing;
(2) Individual model reconstruction;
(3) Non-rigid registration;
(4) Site model construction;
(5) Probabilistic atlas development.

We construct the master model from 90 surgical specimens. We propose an enhanced self-organizing scheme to decompose a set of object contours, representing multi-foci tumors, into localized tumor elements. We apply a mixture of Principal-Axis Registration (mPAR) scheme to align individual prostate models into the site model. Based on accurately mapped tumor distribution, a standard finite normal mixture (SFNM) is used to model volumetric cancer probability density, whose parameters are estimated using K-means and/or Expectation-Maximization (EM) algorithms and the Minimum Description Length (MDL) criterion. We report our evaluation and pilot studies on the effectiveness of the method and its application to optimizing needle biopsy strategies.

## 2. Method

In this section, we describe the major methodological principles and development effort. Three-dimensional digital and optical imaging transforms serial slices of surgical specimen into a computer-synthesized display that facilitates visualization of underlying spatial relationships. Given this digital information, the development of improved computer graphics and visualization has made it possible to study organs and disease patterns in locations that have previously been difficult to evaluate quantitatively.

*2.1 Raw data collection and pre-processing*

All the raw data sets are supplied by the experienced pathologists with computer-aided method to digitize the cross-sectional sequences of real prostatectomy specimens removed due to prostate cancer. We have digitized the cross-sectional sequences of 200 whole mount prostatectomy specimens removed due to prostate cancer provided by the AFIP. In each case the areas of localized tumor have been delineated by an experienced pathologist using computer-aided methods. All the raw data sets need to be pre-processed including data format converting, splitting a group of tumor contours into tumor elements, etc.

In this project, all the raw data representing prostate structures and tumors are given in the format of object contours outlined by the experienced pathologists using optically-imaged surgical specimens and computer-aided methods. The contours of prostate structures have been classified into anatomic objects by the data provider already. Specifically, the contours of prostate capsule are given as class 1, the contours of seminal vesicles are given as class 3, the contours of urethra are given as class 6, and other irrelevant objects are not given a class number. All the tumor contours are given together, as class 5, without any additional information on their multi-foci nature.

*2.2 Individual model reconstruction*

For individual model reconstruction and cancer analysis, there is an urgent need in our research to decompose class 5, a group of contours representing multifoci tumors, into localized tumor elements by a self-organizing method. The decomposition is based on the following assumptions: a tumor contour in level K can only be linked to the contours at the adjacent level $K+1$ or $K-1$. Elementary matching is a 1-to-$n$ ($n >= 1$ is an integer) matching in which a relatively bigger contour at $K$ level can be linked with $n$ smaller contours at $K+1$ or $K-1$ level. Then, there may be a tree-type matching at the two adjacent levels. If $n = 1$, the matched tumor element is a column. If $n \neq 1$, the matched tumor element is a tree or mesh. For a given contour, we use three criteria to search for the most suitable contours at the adjacent levels: maximum area overlap; minimum center distance; and shape similarity (e.g., correlation coefficient).

We denote source contour by $C_{source}$, candidate contour by $C_{candidate}$, and matched contour by $C_{matched}$, and separating parameter by $d_{separating}$ that depends on the tumor shape trend and property of tissue. Shape trend can be estimated by the points of $C_{matched}$. The points are selected from the nearest side of the $C_{source}$ to the $C_{candidate}$ center. The selected points form a curve which can be extended to the level of $C_{candidate}$ in many ways (e.g., a polynomial curve fitting). Based on contour center distance, we can decide whether a $C_{candidate}$ is a $C_{matched}$. Moreover, when the area overlap between $C_{candidate}$ and $C_{source}$ is significant (e.g., >55%), then $C_{scan}$ $C_{candidate}$ is a $C_{matched}$. Furthermore, due to the continuity of organ growth, shape similarity can be used to decided whether a $C_{candidate}$ is a $C_{matched}$.

When $n = 1$, we use area overlap to first select $C_{candidate}$, we then use shape similarity to find $C_{matched}$ among the candidates. If there is no candidate, we use center distance to select $C_{candidate}$. If no $C_{matched}$ is found, then $C_{source}$ is the terminal of an element. When $n \neq 1$, we first use contour center distance to eliminate the contour(s) that cannot be a $C_{candidate}$. We then use shape similarity to each of the remaining $C_{candidate}$, and select the one with the best shape fit.

From 200 surgical specimens, each case consists of 10-14 slices with 4 µm sections at 2.5 mm intervals, and was digitized at a resolution of 1500 dots per inch. Contour extraction was performed by a pathologist followed by a semi-automatic contour refining algorithm using a snake model. The regions of interest (ROI) include the prostate capsule, urethra, seminal vesicles, ejaculatory ducts, surgical margin, prostate carcinoma, and areas of prostatic intraepitrelial neoplasia. For accurate object reconstruction [5], contour interpolation was performed to fill the gaps between one start and one goal contours. Let $\vec{W}^k(i)$ be an intermediate contour, instead of using linear or shape-based interpolation, we developed a 3-D elastic contour model to compute a 3-D force field between adjacent slices thus enabling a "pulling and pushing" metaphor to move the starting contour gradually to the final contour:

$$\vec{W}^k(i) = \vec{W}^{k-1}(i) + \overrightarrow{DS}^{k-1}(x_i^{k-1}, y_i^{k-1}) \qquad (1)$$

where $\overrightarrow{DS}$ is the bilateral force field vector. The nonlinearity characteristics of the elastic contour model permit a meaningful interpolation result yielding a high quality representation of the smoothness nature of the object surface. Reconstruction of an object requires the formation of 3-D surfaces between the contours of successive 2-D slices. Instead of connecting the contours by planar triangle elements where the reconstructed surfaces are usually coarse and static, we developed a physical-based deformable surface model involving two major operations: (1) triangulated patches were tiled between adjacent contours with a criterion of minimizing the surface area, and (2) tiled triangulated patches were refined by using a deformable surface-spine model. Let $v(s,r)$ be the parameterized surface, the associated energy $\mathcal{E}(v)$ can be given [5]:

$$\varepsilon = \int_\Omega \{w_{10}\left\|\frac{\partial v}{\partial s}\right\|^2 + w_{01}\left\|\frac{\partial v}{\partial r}\right\|^2 + 2w_{11}\left\|\frac{\partial^2 v}{\partial s \partial r}\right\|^2 \\ + w_{20}\left\|\frac{\partial^2 v}{\partial s^2}\right\|^2 + w_{02}\left\|\frac{\partial^2 v}{\partial r^2}\right\|^2 + P(v)\}dsdr \qquad (2)$$

where $P(v)$ is the potential of the external forces, and the internal forces are controlled by the coefficients of elasticity ($w_{10}$, $w_{01}$), rigidity ($w_{20}$, $w_{02}$), and twist resistance ($w_{11}$). The surface formation is governed by a second order partial differential equation and is accomplished when the energy of the deformable surface model reaches its minimum. The nonlinear property of the deformable surface model will greatly improve the consistency of the reconstructed complex surface. Using advanced object-oriented 3-D graphics toolkits, interactive visualization is achieved and applied to computerized biopsy simulation. Through efficient picking and surface rendering capabilities, the system allows a user to manipulate simulated needles in the rendered surfaces, the position, orientation and depth of the simulated needle can be specified and recorded.

*2.3 Non-rigid registration*

The estimation of transformational geometry from two point sets is an essential step to medical imaging and computer vision [4]. The task is to recover a matrix representation requiring a set of correspondence matches between features in the two coordinate system. Assume two point sets $\{p_{iA}\}$ and $\{p_{iB}\}$; $i = 1, 2, ... N$ are related by

$$p_{iB} = Rp_{iA} + T + N_i \qquad (3)$$

where R is a rotation matrix, T is a translation vector, and $N_i$ is a noise vector. Given $\{p_{iA}\}$ and $\{p_{iB}\}$, Arun *et al*. present an algorithm for finding the least-squares solution of R and T, which is based on the decoupling of translation and rotation and the singular value decomposition of a 3×3 cross-covariance matrix [4].

The major limitation of the present method is twofold: 1) while feature matching methods can give quite accurate solutions, obtaining correct correspondences of features is a hard problem, especially in the cases of images acquired using different modalities or taken over a period of time and 2) a rigidity assumption is heuristically imposed, leading to the incapability of handling situations with non-rigid deformations. One popular method that does not require correspondences is the principal axes registration (PAR), which is based on the relatively stable geometric properties of image features, i.e., the geometric information contained in these stable image features is often sufficient to determine the transformation between images.

We first discuss the optimality of PAR in a maximum likelihood (ML) sense. The novel feature is to align two point sets without needing to establish explicit point correspondences. We then propose a somewhat different approach for recovering transformational geometry of non-rigid deformations. That is, rather than using a single transformation matrix which gives rise to a large registration error, we attempt to use a mixture of principal axes registrations (mPAR), whose parameters are estimated by minimizing the relative entropy between the two point distributions and using the expectation-maximization algorithm. We demonstrate the principle of the method for both rigid and non-rigid image registration cases.

As suggested by information theory, we note that the control point sets in two images can be considered as two separate realizations of the same random source. Therefore, we do not need to establish point correspondences to extract the transformation matrix. In other words, if we denote $P_{\{p_i\}}$ by the distribution of the control point set in an image, we have the simple relationship

$$P_{\{p_{jB}\}} = P_{\{Rp_{iA}+T\}} + v \qquad (4)$$

where $v$ is the noise component (caused by misalignment). The probability distributions can be computed independently on each image without any need to establish feature correspondences, and given the two distributions of the control point sets in the two images, we can recover the transformation matrix in a simple fashion, as we now describe.

From observation of the distributions, we can estimate R and T by minimizing the relative entropy (Kullback-Leibler distance) between $P_{\{p_{jB}\}}$ and $P_{\{Rp_{iA}+T\}}$, i.e.,

$$\arg \min_{R,T} D\left(P_{\{p_{jB}\}} \| P_{\{Rp_{iA}+T\}}\right) \tag{5}$$

where D denotes the relative entropy measure. We have previously shown the relationship between the negative log joint likelihood and the relative entropy as

$$-\frac{1}{N_B} \log \mathcal{L}\left(P_{\{Rp_{iA}+T\}}(p_{jB})\right) = H\left(P_{\{p_{jB}\}}\right) + D\left(P_{\{p_{jB}\}} \| P_{\{Rp_{iA}+T\}}\right) \tag{6}$$

where $H$ denotes the entropy measure. Thus, minimizing $D\left(P_{\{p_{jB}\}} \| P_{\{Rp_{iA}+T\}}\right)$ is equivalent to maximizing $\log \mathcal{L}\left(P_{\{Rp_{iA}+T\}}(p_{jB})\right)$. Following the same strategy to decouple translation and rotation, we can define a new data point by $q_{iA} = p_{iA} - p_A^0$ and $q_{iB} = p_{iB} - p_B^0$, where $p_A^0$ and $p_B^0$ are the centroids $\{p_{iA}\}$ of $\{p_{iB}\}$ and, respectively. Then the ML estimator of R is defined by

$$\arg \min_R \log \mathcal{L}\left(P_{\{Rp_{iA}+T\}}(q_{jB})\right) \tag{7}$$

and $T = p_B^0 - R p_A^0$

In the case of principal axes technique, we assume a Gaussian model for $P_{\{q_{iA}\}}$ and $P_{\{q_{jB}\}}$. Therefore,

$$\log \mathcal{L}\left(\frac{N_A^{1/2}}{(2\pi)^{3/2} |RC_A R^t|^{1/2}} \times \exp\left(-\frac{1}{2} q_{jB}^t (RC_A R^t)^{-1} q_{jB}\right)\right)$$

$$= \log \frac{N_A^{1/2} N_B}{(2\pi)^{3/2}} - \frac{1}{2} \log |RC_A R^t| - \frac{1}{2N_B} \sum_{j=1}^{N_B} q_{jB}^t (RC_A R^t)^{-1} q_{jB} \tag{8}$$

where the superscript $t$ denotes matrix transposition, $C_0$ denotes the auto-covariance matrix

$$C_A \triangleq \frac{1}{N_A} \sum_{i=1}^{N_A} q_{iA} q_{iA}^t \quad \text{or} \quad C_B \triangleq \frac{1}{N_B} \sum_{i=1}^{N_B} q_{iB} q_{iB}^t \tag{9}$$

and $N_A$ and $N_A$ are the sizes of the point sets $\{q_{iA}\}$ and $\{q_{jB}\}$ respectively. By taking the derivative of (8) with respect to and setting it equal to zero, we have the ML equation (see hints in Appendix)

$$C_B = RC_A R^t \tag{10}$$

Now let the eigenvalue decompositions of $C_A$ and $C_B$ be

$$C_A = U_A \Lambda_A U_A^t, \quad C_B = U_B \Lambda_B U_B^t \tag{11}$$

where $U_A$ and $U_B$ are $3 \times 3$ orthonormal matrices and $\Lambda_A$ and $\Lambda_B$ are $3 \times 3$ diagonal matrices with nonnegative elements. Note that the transformation U consists of the orthonormal set of eigenvectors of C, and matrix $\Lambda$ contains eigenvalues $\lambda_m$ of C for $m = 1,2,3$. Then, we assign

$$R = U_B K U_A^t \tag{12}$$

where K is a $3 \times 3$ diagonal matrix with element $k_m = \sqrt{\lambda_{mB}/\lambda_{mA}}$, the right side of ML (10) becomes $RC_A R^t = U_B K U_A^t U_A \Lambda_A U_A^t U_A K U_B^t = U_B \Lambda_B U_B^t$ which equals exactly the left side of ML (10). Thus, among all $3 \times 3$ orthonormal matrices, R defined by (12) that also includes a scaling matrix *K*, maximizes the joint log likelihood in (8). So far, we have verified the optimality of PAR techniques.

However, because of its global linearity, the application of PAR is necessarily somewhat limited. An alternative paradigm is to model a multimodal control point set with a collection of local linear models. The method is a two-stage procedure: a soft partitioning of the data set followed by estimation of the principal axes within each partition. Recently there has been considerable success in using standard finite normal mixture(SFNM) to model the distribution of a multimodal data set, and the association of a SFNM distribution with PAR offers the possibility of being able to register two images through a mixture of probabilistic principal axes transformations [4].

Assume that there are $K_0$ control point clusters, where each control point cluster defines a transformation $\{R_k, T_k\}$. Thus for a pixel $p_{nA}$, its new locations, corresponding to each of the transformations, are $p_{nk} = R_k p_{nA} + T_k$ for $k = 1, \ldots, K_0$. Further assume that the control point set defines a SFNM distribution

$$f(P_i) = \sum_{k=1}^{K_0} \alpha_k g(p_i | \mu_k, C_k) \qquad (13)$$

where is the Gaussian kernel with mean vector $\mu_k$ and auto-covariance matrix $C_k$, and $\alpha_k$ is the mixing factor which is proportional to the number of control points in cluster *k*. For each of the control point sets $\{p_{iA}\}$ and $\{p_{iB}\}$, the mixture is fit using the expectation-maximization (EM) algorithm [6]. The *E* step involves assigning to the linear models contributions from the control points; the *M* step involves re-estimating the parameters of the linear models in the light of this assignment.

*E-Step*

$$z_{ik}^{(l)} = \frac{\alpha_k^{(l)} g(p_i | \mu_k^{(l)}, C_k^{(l)})}{f\left(p_i | \pi_k^{(l)}, \mu_k^{(l)}, C_k^{(l)}\right)}, \qquad (14)$$

*M-Step*

$$\alpha_k^{(l+1)} = \frac{1}{N} \sum_{i=1}^{N} z_{ik}^{(l)} \qquad (15)$$

$$\mu_k^{(l+1)} = \frac{\sum_{i=1}^{N} z_{ik}^{(l)} p_i}{\sum_{i=1}^{N} z_{ik}^{(l)}} \qquad (16)$$

$$C_k^{(l+1)} = \frac{\sum_{i=1}^{N} z_{ik}^{(l)} (p_i - \mu_k^{(l)})(p_i - \mu_k^{(l)})^t}{\sum_{i=1}^{N} z_{ik}^{(l)}} \qquad (17)$$

For each complete cycle of the algorithm, we first use the "old" set of parameter values to determine the posterior probabilities $z_{ik}^{(l)}$ using (14). These posterior probabilities are then used to obtain "new" values $\alpha_k^{(l+1)}, \mu_k^{(l+1)}, C_k^{(l+1)}$ and using (15)–(17). The algorithm cycles back and forth until the value of relative entropy between the data histogram and mixture model $D\left(P_{\{p_i\}} \| f(p_i)\right)$ reaches its saturation point, for $\{p_{iA}\}$ and $\{p_{iB}\}$, respectively. Our experience indicates that 20 iterations should be sufficient to reach such point, although the number of iterations may vary from case to case occasionally.

Thus the statistical membership of pixel $p_{nA}$ belonging to each of the control (point) clusters can be derived by

$$z_{nk} = P(R_k, T_k | p_{nA}) = \frac{\alpha_{kA} g(p_{nA} | \mu_{kA}, C_{ka})}{f(p_{nA})} \qquad (18)$$

i.e., the posterior probability of $\{R_k, T_k\}$ given $p_{nA}$. We can define the mPAR transformation as [4]

$$\begin{aligned} p_n &= \sum_{k=1}^{K_0} z_{nk} p_{nk} \\ &= \sum_{k=1}^{K_0} \frac{\alpha_{kA} g(p_{nA}|\mu_{kA}, C_{kA})}{f(p_{nA})} (R_k p_{nA} + T_k) \end{aligned} \quad (19)$$

where $\{R_k, T_k\}$ is determined based on $\{(\mu_{kA}, C_{kB}), (\mu_{kB}, C_{kB})\}$ that we have estimated in the previous step using the EM algorithm. Note that now we do need the correspondences between the two control (point) clusters for each $k$. These correspondences may be found, after a global PAR is initially performed, by using a site model approach or a dual-step EM algorithm to unify the tasks of estimating transformation geometry and identifying cluster-correspondence matches. This philosophy for recovering transformational geometry of the nonrigid deformations is similar in spirit to the modular networks in neural computation, under which the relative entropy between the two point sets reaches its minimum

$$\arg\min_{R_k, T_k} D\left( P_{\{p_{jB}\}} \middle\| P_{\left\{\sum_{k=1}^{K_0} z_{ik}(R_k p_{iA} + T_k)\right\}} \right) \quad (20)$$

both globally and locally [4].

*2.4 Site model construction*

The development of statistical modeling and information visualization of localized prostate cancer will require an accurate graphical matching of individual surgical specimens, and volumetric visualization of probability mixture distribution. Based on 200 surgical specimens of the prostates, we have developed a surface reconstruction technique to interactively visualize the clinically significant objects of interest such as the prostate capsule, urethra, seminal vesicles, ejaculatory ducts and the different carcinomas, for each of these cases.

Site model construction is performed by a coupled dynamic deformation system [7]. The axis of the surface from new contours connects all the surface patches to the spine through expansion/compression forces radiating from the spine while the spine itself is also confined to the surfaces. The dynamics is governed by the second-order partial differential equations from Lagrangian mechanics so that final shapes and relationships of the surface and spine are achieved with a minimum energy dynamic deformation. Let the strain energies of surface ($\mathcal{E}_{surface}$) and spine ($\mathcal{E}_{spine}$) be the sum of controlled stretching and bending energies, where $\mathcal{E}_{surface}$ is the thin-plate under tension variational spline and $\mathcal{E}_{spine}$ is a weighted sum of the tension along the spine (stretching energy) and the controlled rigidity (bending energy), the non-rigid motion in response to an extrinsic force f(x) follows the continuum mechanical equation [7]

$$\mu \frac{\partial^2 \mathbf{x}}{\partial t} + \gamma \frac{\partial \mathbf{x}}{\partial t} + \frac{\delta \varepsilon(\mathbf{x})}{\delta \mathbf{x}} = f(\mathbf{x}), \quad (21)$$

where μ is the mass density function, γ is the viscosity function, and $\frac{\delta \varepsilon(\mathbf{x})}{\delta \mathbf{x}}$ is the variational derivative of $\mathcal{E}$ representing the internal elastic force. After principal-axes based initial registration, external forces are introduced to both the surfaces and spines to be reconstructed or matched, such as $f_a$ as a function of the difference between the spine and the axis of the surface, radial force $f_b$, and inflation or deflation force $f_c$. The control over expansion and contraction of the surface around the spine, realized by summing the coupled forces, leads to the following dynamic system [7]:

$$\mu \frac{\partial^2 \mathbf{x}_{surface}}{\partial t^2} + \gamma \frac{\partial \mathbf{x}_{surface}}{\partial t} + \frac{\delta \varepsilon_{surface}}{\delta \mathbf{x}} = \mathbf{f}^{ext}_{surface} + \mathbf{f}^{a}_{surface} + \mathbf{f}^{b}_{surface} + \mathbf{f}^{c}_{surface} \qquad (22)$$

$$\mu \frac{\partial^2 \mathbf{x}_{spine}}{\partial t^2} + \gamma \frac{\partial \mathbf{x}_{spine}}{\partial t} + \frac{\delta \varepsilon_{spine}}{\delta \mathbf{x}} = \mathbf{f}^{ext}_{spine} + \mathbf{f}^{a}_{spine} \qquad (23)$$

where x$_{surface}$ is the vector of the coordinates of a point on the surface, and x$_{spjne}$ the vector of the coordinates of a point on spine in R$^3$. The $\mathbf{f}^{ext}_{surface}$ the external force applied on the surface and $\mathbf{f}^{ext}_{surface}$ the external force applied on the spine. We developed a finite element method (FEM) using 9 degree-of-freedoms (dofs) triangular elements for x$_{surface}$ and 4 dofs spine elements for x$_{spjne}$, where the dofs at each node correspond to its position and parametric tangent(s). The combinational capability of deformable FEM on the basis functions $N_i$ (so called shape functions) provides a continuous surface (or spine) representation of a large range of topological shapes. Furthermore, started with one reconstructed surface, each vertex of the triangular element is analyzed to determine the appropriate external forces from its closely matched point on the second surface. With all 3-D force fields determined, we apply the Lagrangian non-rigid motion equations to dynamically move the starting surface to the target, the second surface. In particular, other surfaces in this dynamic system can be transformed to their final positions with nonlinearly deformed shapes in consistency by applying corresponding force fields (Gaussian weighted sums of the external forces on the starting surface) onto these surfaces. The iterative process of matching is terminated and accomplished by reaching the minimum energy dynamic deformation [7].

The deformable energy of surface x(u,v,t) can be defined by

$$\varepsilon_{surface} = \int_2^1 \int_2^1 \left( w_{10} \left|\frac{\partial x}{\partial u}\right|^2 + 2w_{11} \left|\frac{\partial x}{\partial u}\right|\left|\frac{\partial x}{\partial v}\right| + w_{01}\left|\frac{\partial x}{\partial v}\right|^2 + w_{20}\left|\frac{\partial^2 x}{\partial u^2}\right|^2 + 2w_{22}\left|\frac{\partial^2 x}{\partial u \partial v}\right|^2 + w_{02}\left|\frac{\partial^2 x}{\partial v^2}\right|^2 \right) du\,dv, \qquad (24)$$

where the weight $w_{10}$, $w_{11}$ and $w_{01}$ control the tensions of the surface, while $w_{20}$, $w_{22}$, and $w_{02}$ control its rigidities(bending energy). The deformable energy of spine x(s,t) is given by

$$\varepsilon_{spine(u,v,t)} = \int_2^1 (w_1 \left|\frac{dx}{ds}\right|^2 + w_2 \left|\frac{d^2 x}{ds^2}\right|^2 )ds. \qquad (25)$$

The weight $w_1$ controls the tension along the spine(stretching energy), while $w_2$ controls its rigidity(bending energy).

To couple the surface with the spine, we enforce $v \equiv s$, which maps the spine coordinate into the coordinate along the length of the surface. Then we connect the spine with surface by introducing following forces on the surface and spine respectively:

$$\mathbf{f}^{a}_{surface}(u,s,t) = -(a/l)(\overline{X}_{surface} - X_{spine})$$
$$\mathbf{f}^{a}_{spine}(s,t) = a(\overline{X}_{surface} - X_{spine}) \qquad (26)$$

Where $a$ controls the strength of the forces; $\overline{X}_{surface}$ is the centroid of the coordinate curve(s=constant) circling the surface and is defined as: $\overline{X}_{surface} = \frac{1}{l}\int_0^1 X_{surface}\left|\frac{\partial x_{surface}}{\partial u}\right|du$, and $l$ is the length of the curve (s=constant). In general, the above forces coerce the spine staying on an axial position of surface. Further, if necessary, we can encourage the surface to be radially symmetric around the spine by introducing the following force:

$$\mathbf{f}^{b}_{surface} = b(\overline{r} - |r|)\hat{r}, \qquad (27)$$

where $b$ controls the strength of the force; $r$ is the radial vector of the surface with respect to the spine as r(u,s)=x$_{surface}$-x$_{spine}$, the unit radial vector is $\hat{r}(u,s) = r/|r|$, and $\overline{r}(s) = \frac{1}{l}\int_0^1 |r|\left|\frac{\partial x_{surface}}{\partial u}\right|du$ is the mean radius of the coordinate curve *s*=constant. Also it is possible to provide control over expansion and contraction of the surface around the spine. This can be realized by introducing the following force:

$$f^c_{surface} = c\hat{r}, \qquad (28)$$

where *c* controls the strength of the expansion or contraction force. The surface will inflate where $c > 0$ and deflate where $c < 0$.

Summing the above coupling forces in the motion equation associated with surface and spine, we obtain the following dynamic system describing the deformable surface-spine model:

$$\mu \frac{\partial^2 x_{surface}}{\partial t^2} + \gamma \frac{\partial x_{surface}}{\partial t} + \frac{\delta \mathcal{E}_{surface}}{\delta x} = f^{total}_{surface},$$
$$\mu \frac{\partial^2 x_{spine}}{\partial t^2} + \gamma \frac{\partial x_{spine}}{\partial t} + \frac{\delta \mathcal{E}_{spine}}{\delta x} = f^{total}_{surface}, \qquad (29)$$

where $f^{total}_{surface} = f^{ext}_{surface} + f^a_{surface} + f^b_{surface} + f^c_{surface}$ and $f^{total}_{spine} = f^{ext}_{spine} + f^a_{spine}$. Note that $f^{total}_{surface}$ is the external force applied on the surface and $f^{ext}_{spine}$ the external force applied on the spine(we assign $f^{ext}_{spine} = 0$ in our implementation) .We surface registration problem, we are interested in matching two surfaces by computing the deformation between them. We define the external forces to reflect the distance between the two surfaces under consideration:

$$f^{ext}_{surface} = C(x_{surface}(u,v)), \qquad (30)$$

where $C(x_{surface}(u,v))$ is the Euclidean distance of each point on the surface to the nearest point on the second surface. The final $x_{surface}$ and $x_{spine}$ are obtained when the energy of the deformable surface-spine reaches its minimum. To solve equation (29) of such a dynamic system, we have developed several force balance strategies to perform 3D model to model warping. We have also extended the surface-spine model to a deformable coupled-surface model where the "spine" is replaced by the coupled "surface" to generate a blended generic model [7].

*2.5 Probabilistic atlas development*

Based on an accurate multi-object and non-rigid registration of tumor distribution, a standard finite normal mixture is applied to model statistics of the cancer volumetric distribution, whose parameters are estimated using K-means algorithm as the initialize sites and using expectation-maximization algorithms, under the information theoretic criteria, to finalize the sites. The development of the biopsy site selection consists of two steps:

(1) Using K-means algorithms to initialize the cluster centers of the cancer volumetric distribution model.

(2) Using EM algorithms to optimize through a probabilistic self-organizing map to achieve a maximum likelihood of cancer detection.

Step 1: K-means algorithms is one of the important issues in pattern classification to find a set of representative vectors for clouds of multimodality data sets. Pattern vectors of *n*-dimensions may be considered as representing points within an *n*-dimensional Euclidean space. One of the most obvious means by which we may establish a measure of similarity among such pattern vectors is by means of their proximity to one another. The K-means algorithm is one of many clustering techniques that share the notion of clustering by minimum distance. For our purposes, we used the K-means algorithm to initialize the cluster centers of the prostate cancer distribution. Given the number of cluster centers of interest, the K-means algorithm can determine the initial locations of the cluster centers of the probability map of cancer distribution.

This parallel method initially takes the number of cluster centers of the interest equal to the final required number of clusters. In this step the final required number of cluster centers be chosen such that the points are mutually farthest apart. Next, it examines each component in the data sets and assigns it to one of the clusters depending on the minimum distance. The centroid's position is recalculated every time a component is added to the cluster and this continues until all the components are converged into the final required number of clusters.

Step 2: the expectation-maximization (EM) algorithm provides an iterative approach to compute maximum likelihood estimates in situations where, the given observations are either incomplete or can be viewed as incomplete [6]. The EM algorithm derives its name from the fact that on each iteration of the algorithm there are two steps, which are the expectation step and the maximization step. The expectation step uses the observed data set of an incomplete data problem and the current value of the parameter vector to manufacture data so as to postulate an augmented or so called complete data set. The maximization step consists of deriving a new estimate of the parameter vector by maximizing the log likelihood function to complete data manufactured in the E-step. Thus, starting from a suitable value for the parameter vector, the E-step and M-step are repeated on an alternate basis until converged.

After we acquired the initial cluster centers from the K-means algorithm, we have applied the EM algorithm to estimate the posterior Bayesian probabilistic class memberships of the data point with respect to each of the local classes. The EM algorithm can accurately and effectively classify the data points into correct classes. By combining these two procedures, it will help us to optimize the prostate needle biopsy site selection through a probabilistic self-organizing map, thus achieving a maximum likelihood of cancer detection.

In order to quantitatively investigate the tumor distribution, volume, and multi-foci in space, a statistical master model of localized prostate cancer will be required to relate individual graphical models to a global probability distribution. Based on 200 reconstructed and registered computer models representing the prostate capsule and internal structures, each of these cases has been automatically aligned together. By labeling the voxels of localized prostate cancer by "1" and the voxels of other internal structures by "0", we generated a 3-D binary map of the prostate that is simply a mutually exclusive random sampling of the underlying spatial probability distribution of cancer occurrence. We have summarized all these binary maps and normalized the result to obtain a 3-D histogram of the cancer distribution that can be modeled by a SFNM:

$$f(x,y,z) = \sum_{k=1}^{K} \pi_k G(x,y,z|\mathbf{\mu}_k, \mathbf{\Sigma}_k), \tag{31}$$

where $\mathbf{\mu}_k$ and $\mathbf{\Sigma}_k$ are the mean vector and covariance matrix of the $k$th component, $\pi_k$ is the global regularization parameter. The inaccuracy in model selection will affect the performances of both data quantification and classification. Using the proposed information theoretic criteria, the structure of the probabilistic modular networks will be optimized following the model selection, which according to a newly developed information theoretic criterion:

$$MCBV(K,\alpha) = -\log(\mathcal{L}(\mathbf{u}|\hat{\mathbf{r}}_{ML}) + \sum_{k=1}^{3K} H(\hat{\mathbf{r}}_{ML})) \tag{32}$$

where $H(\hat{\mathbf{r}}_{ML})$ is the entropy of the maximum likelihood (ML) parameter estimate $\hat{\mathbf{r}}_{ML}$ obtained through a probabilistic neural network [6]. As a result, MCBV criterion selects a statistical master model with $K_0$ centers shape such that the estimation of the spatial distribution achieves the minimum bias and variance at the same time.

## 3. Experimental results and discussion

In this section, we present results using the principal axes technique and dynamic deformation system based approach we introduced to match individual reconstructed prostate graphical models and fuse them together. After generating a 3-D histogram of localized prostate cancer based on registered samples, we concentrate on statistical modeling of the spatial probability of prostate cancer occurrence, which presents a great challenge to information visualization scheme, because of its stochastic and complex structure.

Fig. 1 shows the major steps of algorithm pipeline, with transparent graphical models reconstructed by the proposed computer algorithm. The regions of interest (ROI) include the prostate capsule, urethra, seminal vesicles, ejaculatory ducts, surgical margin, prostate carcinoma, and areas of prostatic intraepitrelial neoplasia. The results are very consistent with the pathologists' visual inspection and judgment. Fig. 2 shows the graphic user interface for deformation based object surface reconstruction, with typical examples. Fig. 3. Shows mPAR based non-rigid registration. The result of initial registration using mPAR is shown as the left-side contour sets. It can be seen that the principal axes of two multi-foci objects have been aligned fairly accurate (right-side contour sets). The two models before registration are shown as middle contour sets. Fig. 4 shows the improved model fusion by thin-plate spline and dynamic deformation in site model construction. Applying dynamic deformation system to the initial state, active object will converges to the targeted object. Fig. 5 shows mapping multi-foci tumors of individual models into the site model. Fig. 6 show the statistical atlas of localized prostate tumors reconstructed from 90 surgical specimens. As we have discussed before, the ultimate goal of case collection and 3-D matching is to create a master model of localized prostate cancer representing a spatial probability distribution. Using the proposed method, we have for the first time estimated the possible distribution that may reveal the important disease patterns of the prostate cancer. It should be noticed that multi-centricity pattern is clearly shown in the model and the spatial distribution is not uniformly random.

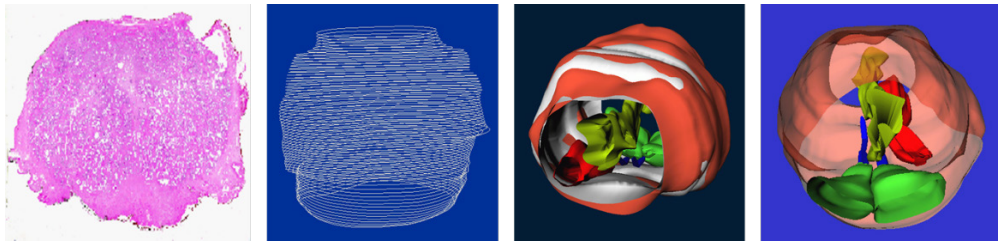

Fig. 1. Major steps and outcomes of individual model construction. (a) Optically-imaged prostate specimen. (b) Object contours outlined by the pathologist(s). (c) Reconstructed graphic model. (d) Transparent visualization of the reconstructed model with internal anatomic structures.

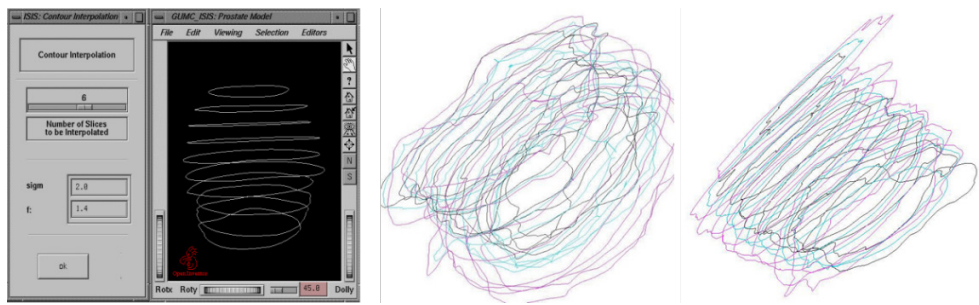

Fig. 2. Individual model reconstruction with user interface. (a) Graphic User Interface. (b) Multi-object deformation based surface reconstruction (case 1: relatively smooth). (c) Multi-object deformation based surface reconstruction (case 2: more challenging).

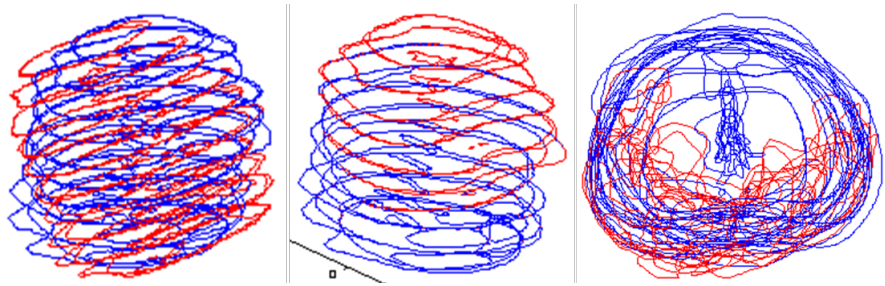

Fig. 3. mPAR based non-rigid registration. (a) Two contour sets after registration. (b) Two contour sets before registration. (c) See-through view of the two aligned contour sets.

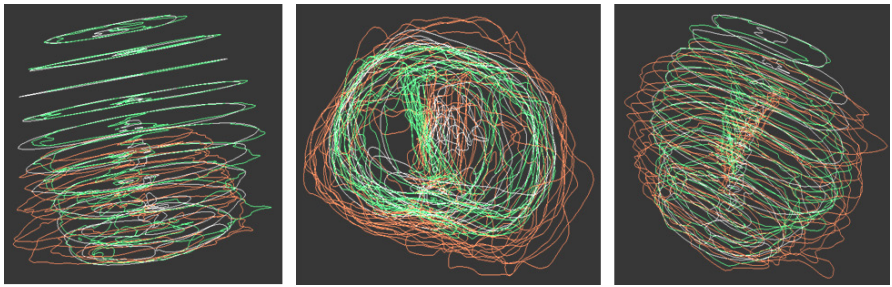

Fig. 4. Thin-plat spline and dynamic deformation based site model construction. (a) Two contour sets before registration. (b) Two contour sets after registration (horizontal view). (c) Two contour sets after registration (vertical view).

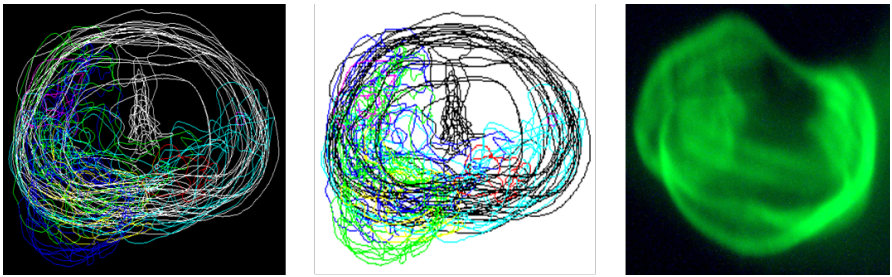

Fig. 5. Mapping of multi-foci localized prostate tumors into the site model. Different colors of the contour sets represent different anatomic structure including tumors.

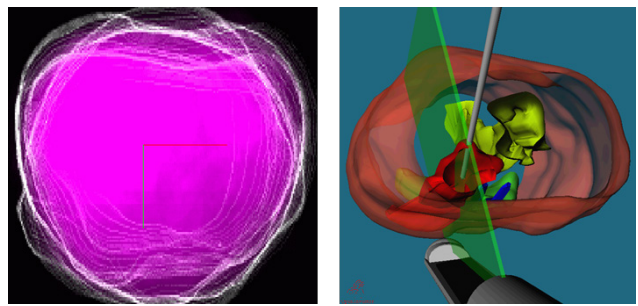

Fig. 6. Statistical atlas of localized prostate tumors reconstructed from 90 surgical specimens.

## 4. Summary

In this paper, a statistically significant master model of localized prostate cancer is developed with pathologically proven surgical specimens to spatially guide specific points in the biopsy technique for a higher rate of prostate cancer detection and the best possible representation of tumor grade and extension. In order to investigate the complex disease pattern including the tumor distribution, volume, and multi-centricity, we created a statistically significant master model of localized prostate cancer by fusing these reconstructed computer models together, followed by a quantitative formulation of the 3-D finite mixture distribution. Based on the reconstructed prostate capsule and internal structures, we have developed a technique to align all surgical specimens through elastic matching. The preliminary results show that a statistical pattern of localized prostate cancer exists, and a better understanding of disease patterns associated with tumor volume, distribution, and multi-foci of prostate carcinoma can be obtained from the computerized master model. While such statistical atlas can serve as the 'prior', the next major step would be to combine this master with the advanced in vivo molecular imaging technologies for developed truly imaging-informed biopsy strategies [8, 9].